\documentclass[
review]{elsarticle
}
\usepackage[T1]{fontenc}
\usepackage{amsmath}
\usepackage{amssymb}
\usepackage{graphicx}
\usepackage{color}
\usepackage[inline]{enumitem}
\usepackage[utf8]{inputenc}
\usepackage{tikz,pgfplots}

\pdfpageheight\paperheight
\pdfpagewidth\paperwidth

\definecolor{byzantine}{rgb}{0.74, 0.2, 0.64}\definecolor{vividviolet}{rgb}{0.62, 0.0, 1.0}
\newcommand{\Mov}[1]{{\color{black}{#1}}}
\newcommand{\Jav}[1]{{\color{black}{#1}}}

\pgfplotsset{compat=1.14}

\begin{document}
\begin{frontmatter}

\title{Is the Cosmological Constant of Topological Origin?}

\author[ad1,ad2,ad3]{M. Le Delliou}
\ead{delliou@lzu.edu.cn,Morgan.LeDelliou.IFT@gmail.com}
\address[ad1]{Institute of Theoretical Physics, School of Physical Science and Technology, Lanzhou University,
No.222, South Tianshui Road, Lanzhou, Gansu 730000, P R China}
\address[ad2]{Instituto de Astrof\'isica e Ci\^encias do Espa\c co, Universidade de Lisboa,
Faculdade de Ci\^encias, Ed.~C8, Campo Grande, 1769-016 Lisboa, Portugal}
\address[ad3]{Instituto de F\'isica Te\'orica, Universidade Estadual Paulista (IFT-UNESP), Rua Dr. Bento Teobaldo Ferraz 271, Bloco 2 - Barra Funda,  CEP 01140-070, S\~ao Paulo, SP, Brazil}
\author[adl]{J. Lorca Espiro}%
\ead{javier.lorca@ufrontera.cl}
\address[adl]{Departamento de Ciencias F\'\i sicas, Facultad de Ingenier\'\i a, Ciencias y Administraci\'on, Universidad de La Frontera, Avda. Francisco Salazar 01145, Casilla 54-D Temuco, Chile.}


\begin{abstract}
The observed value of the cosmological constant poses large theoretical problems. We find that topology of the Universe provides a natural source for it. Restricting dynamically an Einstein-Cartan gravity to General Relativity in our observed Universe allows topological invariants to induce an effective cosmological constant from dynamical quintessence-like topological fields. Its evaluation through the boundary of black holes yields a range compatible with the observed value, with uncertainty of three orders of magnitude. In turns, it provides a measurement of the Universe's isoperimetric constant. \Jav{As in other dark energy studies, quantum vacuum energy corrections are left for quantum gravity studies to explain away.}
\end{abstract}

%
%

\end{frontmatter}
\section{Introduction}
The cosmological constant has reappeared at the turn of the century in the toolbox of cosmology with the discovery of fainter than expected distant type Ia supernovae \cite{Perlmutter:1997zf,Riess:1998cb,Perlmutter:1998np} interpreted as cosmic expansion acceleration. This acceleration was confirmed by the combination of cosmic microwave background radiation, clusters and baryon acoustic oscillation observations \cite{Hinshaw:2012aka,Bennett:2012zja,Aghanim:2012vda,Ade:2013sjv}.

Its simple interpretation as quantum vacuum energy clashes with one of the largest discrepancy of physics: its observed value from the previous cosmic geometry tools yields a value for $\Lambda$ that contradicts the simple evaluation that it should reach order of the Planck scale \cite{Martin}. This has been coined the fine tuning problem \citep{Weinberg:1988cp,Weinberg:2008zzc}. Quantum vacuum is ruled by the Planck scale at early time which then should set its initial value, but since $\Lambda$ is a constant, this is the value expected nowadays, far from the hundred order of magnitude lower observed. 

In addition, this non-varying value gives rise to the coincidence problem \citep{Amendola:1999er,TocchiniValentini:2001ty,Zimdahl:2001ar,Zimdahl:2002zb} as it also yields a very recent epoch for its emergence as cosmic dominant component: its energy density, set at the initial universe conditions, is unnaturally close to that of present matter and therefore poses some questions \cite{Frieman:2008sn,Li:2012dt}. For a review of the questions posed by the cosmological constant, refer to \cite{Martin}.

The geometrical approach to gravity through curved spacetime by general relativity (GR) can be generalised to include the possibility of its torsion in Einstein-Cartan (EC) theories \cite{Dona,Giulini,nakahara,hatcher}, where GR appears as the torsion-less limit, while the curvature-less limit yields the Teleparallel Equivalent to GR case \cite{Deandrade,Baez,Arcos}. This work will focus on the GR limit of Einstein-Cartan theories, further restricted by adding the characteristic classes consistent with the topology of the space-time manifold $\mathcal{M}$ \cite{donaldson,Sengupta,Kaul}, i.e. the Euler class $e (T \mathcal{M}) := C_E$, the Pontryagin class $p_1 (T \mathcal{M}):= C_P$ and the Nieh-Yan class (of the Chern-type) $c_2 (T \mathcal{M}):= C_N$ \cite{nash,nakahara,donaldson}. In order for these topological terms to be non trivial, we are lead to assume that the space-time manifold has a boundary $\partial \mathcal{M} \neq \emptyset$. 
We found it is composed by the hyper-surfaces that define the horizons of black holes in the Universe. 

The action is then modified so as to include the appropriate boundary counter terms \cite{Baekler,Dyer} and such as the field equations are well-posed. The entire procedure will result in the appearance of a \textit{cosmological functional}~$\tilde{\lambda}$. The dynamical restriction of such torsional theory in $\mathcal{M}$ to GR conditions, in a region $\mathcal{N}\subset\mathcal{M}$ that contains our observed Universe, is achieved by means of the Vielbein-Einstein-Palatini (VEP) formalism and by dynamical systems stability considerations. This allows for the collateral topological \textit{effective cosmological constant}~$\Lambda$ which is then identified with the observed value. 
Given that our result is mostly of topological origin, it should not be affected by quantum corrections, and specifically by quantum fluctuations in the vacuum. \Jav{However, we will not address this issue in the present paper since, as in other dark energy studies, we start from the \Mov{expectation }
that quantum vacuum energy corrections \Mov{should be explained away by quantum gravity considerations }
in cosmological settings.
}

In Sec.~\ref{sec:VEPtoGRregions}, we present the inclusion of topological invariants in the Vielbein-Einstein-Palatini action and how their dynamics can recover a GR behaviour. The emergence of an effective topological cosmological constant is presented in Sec.~\ref{sec:EffTCC}, while its evaluation using black holes occupy Sec.~\ref{sec:BHcc}, before discussing our conclusions in Sec.~\ref{sec:Conclusion}.

\section{Modified VEP restricted to GR regions}\label{sec:VEPtoGRregions}

We focus on the family of actions defined over the manifold $\mathcal{M}$ with boundary $\partial \mathcal{M} \neq \emptyset$, having the following terms: 
\begin{align}\label{Bodyaction}
S & := S_{G} \left[e^c, \omega^a_{\;\;b} , \tilde{\lambda} \right] + S_{T} \left[e^c, \omega^a_{\;\;b}, \varphi_j \right] +  S_M \;\;\; ,
\end{align}
where $\omega^a_{\;\;b}$ is the total connection \cite[see][Ch 7.10 - 7.15]{nash}, $e^c$ the vierbein frame \cite[see][Sec.~7.8]{nakahara} and $\tilde{\lambda}$ is the, thus far unrestricted, \textit{cosmological functional}, with:
\begin{align}
\label{EHactionmod} S_{G} \left[e^c, \omega^a_{\;\;b} , \tilde{\lambda} \right] & = \int\limits_{\mathcal{M}} \frac{1}{\kappa } \left( {R}^{\left( * \right)}_{ab} \wedge \hat{\Sigma}^{ab} - \tilde{\lambda} \, {d \mu} \right) + \int\limits_{ \partial \mathcal{M}} i_n \left( \frac{\tilde{\lambda} \, d \mu}{\kappa} \right) \, ,
\end{align}
is the pure gravitational VEP action \cite{Dadhich} plus a boundary term \textit{\`{a} la} York-Gibbons-Hawkins induced by the inclusion in the boundary operator $i_n ( \cdot ) $, being $\kappa$ the scaled gravitational constant, $R_{ab}$ is the curvature $2$-form \cite[see][Sec.~10.3]{nakahara}, $\hat{\Sigma}^{ab}:= \frac{1}{2} e^a \wedge e^b$ is the Palatini $2$-form\footnote{We are using the notation $A^{\left( * \right)}_{ab} : = \frac{1}{2} \epsilon_{abcd} A^{cd}$ for the Lie dual acting over any $A^{cd} \in \Omega \left( \mathcal{M} \right)$ with two spin indices $c,d$.} and $d \mu := \frac{1}{4!} \epsilon_{abcd} e^a \wedge e^b \wedge e^c \wedge e^d$ is the canonical volume measure. The second term in Eq.~ (\ref{Bodyaction}) is
\begin{align}
\label{Stop}  S_{T} \left[e^c, \omega^a_{\;\;b}, \varphi_j \right] & := - \int\limits_{\mathcal{M}} \frac{i}{\kappa} \left( \varphi_j C_j \right) + \int\limits_{\partial \mathcal{M}} \frac{i}{\kappa} i_n \left( \varphi_j C_j \right)
\end{align} 
which we have called the \textit{topological action}. The coupling zero forms $\varphi_j$ ($j=E,P,N$ explained below) can be interpreted as Lagrange's multipliers for the characteristic classes $C_j$ \cite[see][Ch. 11]{nakahara} \cite[and][Secs.~7.22 - 7.26]{nash}:
\begin{align}
\label{Pontryagin1} \small C_{P}  & = \frac{1}{8 \pi^2} R^a_{\;\;b} \wedge R^b_{\;\;a} \;\;, &&\text{(Pontryagin)} \\
\label{Euler1} \small C_{E} & = \frac{1}{8 \pi^2} R^{ab} \wedge R^{\left( * \right)}_{ba} \;\;,  &&\text{(Euler)} \\
\label{NiehYan1}\small C_N & = T^a \wedge T_a - R_{ab} \wedge \Sigma^{ab} \;\;,  &&\text{(Nieh-Yan)}
\end{align}
again with the appropriate boundary counter terms to render the total action well posed. Another way of motivating $S_T$ is to interpret these terms from the Schwinger's perspective as \textit{current-like} terms with topological sources for the total action. Finally, we take:
\begin{align}\label{mataction}
S_M := S_M \left[ \Psi \right] & = - \int\limits_{\mathcal{M}} \frac{2}{\kappa} \mathfrak{L}_M \left[ \Psi \right] + \int\limits_{\partial \mathcal{M}} \frac{2}{\kappa} i_n\left( \mathfrak{L}_M \left[ \Psi \right] \right) \\
\nonumber \text{with} \quad \mathfrak{L}_M \left[ \Psi \right] & := \mathfrak{Re} \left\{ \bar{\psi} \gamma^a D_{\omega} \psi \right\} \wedge \star e^{\flat}_a - V \left( \Psi \right)  d \mu \;\; ,
\end{align}
as a Dirac action with appropriate boundary counter terms, where $\Psi := \left\{ \psi , \bar{\psi}  \right\}$ stands for mass-less spinor fields, where $\star$ stands for the Hodge Dual operation \cite[see][Sec.~7.9]{nakahara}\footnote{The hodge dual operator is defined over the vierbein basis as $\star \left( e^{a_1} \wedge \cdots \wedge e^{a_n} \right) := \frac{1}{\left( 4 - n\right)!} \epsilon^{a_1 \cdots a_n}_{\;\;\;\;\;\; \;\;\;\;\;a_{n+1} \cdots a_4} e^{a_{n+1}} \wedge \cdots e^4$ and it extends to the entire space $\Omega \left( \mathcal{M} \right)$ by linearity.}.

\subsection{Field equations}

The variation of (\ref{Bodyaction}) gives the following field equations (for $\delta e^a$, $\delta \omega_{ab}$, $\delta \varphi_j$, $\delta \bar{\psi}$ and $\delta \psi$ , respectively)\footnote{Originally the field equations suppose the presence of a fermion potential $V \left( \Psi \right)$. However, since this potential is not involved in the final calculation, we prefer to omit it out of clarity.}:
\begin{align}
\label{fe1} & 0 = \left[ R^{\left( *\right)}_{ab} - \frac{2 \lambda}{3} \left| \psi \right|^2_{\psi} \hat{\Sigma}^{(*)}_{ab} \right] \wedge e^{b} + i d \varphi_{N} \wedge T_{a} - \frac{ \delta S_{M}}{\delta e^a}  \;\; , \\
\label{fe2} & 0 = d_{\omega} \hat{\Sigma}^{(*)}_{ab} + i 2 d \hat{\varphi}_P \wedge R_{ab} - i 2 d \hat{\varphi}_E \wedge R^{\left( * \right)}_{ab} - i d \varphi_{N} \wedge \hat{\Sigma}^{\flat}_{ab} + \frac{\delta S_M}{\delta \omega^{ab} } \;\;\;, \\ 
\label{fe3} & 0  = - i \, \delta_{\varphi_j} \lambda \, \left| \psi \right|^2_{\psi} d \mu - C_j \;\;\;\;\;\; , \;\;\;\;\;\; j = \left\{ E, P, N \right\} \;\; , \\
\label{fe4} & 0 = \gamma^a D_{\omega} \psi - \frac{\lambda}{4} \psi \, e^a \;\;\;\;\; , \;\;\;\; 0 = D_{\omega} \bar{\psi} \gamma^a + \frac{\lambda}{4} \bar{\psi} \, e^a \;\;\; ,
\end{align}
where $T_a$ is the torsion $2$-form, $d_{\omega}$ stands for the exterior covariant derivative $\delta_{\varphi_j}$ is the Euler-Lagrange derivative and $\flat$ is the musical isomorphism \cite[see][Ch.~3]{lee1997riemannian} between vectors and $1$-forms\footnote{Explicitly, given a vector field $X = X^i e_i$ its flat is the $1$-form $X^{\flat }:= X_j e^j.$}. In order to obtain the pair of equations (\ref{fe4}) we have explicitly considered the cosmological functional $\tilde{\lambda}$ to be the Yukawa-type interaction:
\begin{align}\label{Yukawa}
 \tilde{\lambda} = \tilde{\lambda} \left[ \varphi , \Psi \right] := \bar{\psi} \lambda \left[ \varphi_j \right] \psi
\end{align}
and we call $\lambda$ the \textit{reduced cosmological functional}, $D_{\omega}$ stands for the gauge covariant derivative and we have defined the normalized couplings $\hat{\varphi}_j := \frac{\varphi_j}{\left( 4 \pi \right)^2}$ for convenience.

We recall that the total connection $1$-form $\omega^{a}_{\;\;b}$ satisfies the splitting:
\begin{align}\label{connectiondecomp}
\omega^{a}_{\;\;b} = \bar \omega^{a}_{\;\;b} + K^{a}_{\;\;b} \;\; \Rightarrow \;\; T^a = K^{a}_{\;\;b} \wedge e^b \;\;,
\end{align}
where $\bar \omega^{a}_{\;\;b}$ is the torsion-less Levi-Civita connection, while $ K^{a}_{\;\;b}$ is the contortion $1$-form. By means of decomposition (\ref{connectiondecomp}), the total curvature $2$-form can then be written as:
\begin{align}\label{curvaturedecomp}
R^a_{\;\;b} = d \omega^a_{\;\;b} + \omega^a_{\;\; c} \wedge \omega^c_{\;\;b} =  \bar{R}^a_{\;\;b} + \Theta^a_{\;\;b}  \;\;,
\end{align}
where $\bar{R}^a_{\;\;b}$ is the torsion-less part of the curvature, formally equivalent to that of a GR curvature $2$-form, while the term:
\begin{align}\label{torcurvB}
\Theta^a_{\;\;b} & := d_{\omega} K^a_{\;\;b} - K^a_{\;\;c} \wedge K^{c}_{\;\;b} \quad ;
\end{align}
concentrates all the contributions from the torsion related quantities.

\subsection{GR restriction}

Focusing on the field equations  (\ref{fe1} - \ref{fe2}) and keeping in mind the decomposition of the curvature (\ref{curvaturedecomp}) and Eqs.~(\ref{fe4}), it appears that the key to find consistent solutions lies in finding a proper contortion $1$-form $K_{ab}$. 
Our approach seeks to 
cancel out the torsion dependent term $T_a$ in Eq. (\ref{fe1}) in order to obtain a form that resembles more that of GR. 
This can be achieved from the torsion part of the curvature $2$-form $R_{ab}$ via choosing an appropriate contortion: 
\begin{align}\label{contortionbody}
K_{ab} = 4i \star \left( d \varphi_N \wedge \hat{\Sigma}^{\flat}_{ab} \right) \;\;\;.
\end{align}
At the same time, we maintain the independence of the GR-like torsion-less part of the curvature from the torsion part in Eq. (\ref{fe2}) by restricting all the couplings $\varphi_j$ to be related by a functional behavior of the form: \begin{align}\label{jacobians}
d \varphi_j := {\varphi_N}^* \, d \phi_j = d \left( \phi_j \circ \varphi_N \right) = \left( \frac{\partial \phi_j}{\partial \varphi_N} \right) d \varphi_N  \;\;\;\; \text{for} \;\;\;\; j = E, P \quad ,
\end{align}
where $\phi_j$ proceeds from the field equations. The dynamical system defined by Eqs.~(\ref{fe1}) and (\ref{fe2}) drives the remaining independent coupling $\varphi_N$ to be a slowly varying function in a region $\mathcal{N}$, rendering it torsion-less and thus GR-like.

Following this, the topological characteristic classes can be calculated via (\ref{fe3}) and matched with their canonical definitions (\ref{Pontryagin1} - \ref{NiehYan1}) to finally restrict the remaining parameters. Imposing dynamical stability we obtain an expression for the reduced cosmological functional $\lambda$ of the form:
\begin{align}\label{deltalambda1}
\lambda = \frac{4 \Lambda}{\left| \psi \right|^2_{\psi}} u_{{\varphi}_N} \;\;\; \text{where} \;\;\;  u_{{\varphi}_N} := \exp \left( - \frac{\sqrt{3}}{4} \left| \varphi_N \right| \right)
\end{align}
where $\Lambda$ is now a constant. On the other hand, when using the set of equations (\ref{fe3}), the reduced cosmological functional $\lambda$ possesses an equivalent form which allows us to recognize $\varphi_N$ as a mass-less scalar matter field. When combining the latter with Eq.~(\ref{deltalambda1}) we see that the kinetic term associated to the zero-form $\varphi_N$ should behave as a Lyapunov function. Leading to the following expression:
\begin{align*}
\left| \frac{d \varphi_N}{4}  \right|^2 & = \Lambda \left\{ 1 - u_{{\varphi}_N} \right\} \;\; ,
\end{align*}
which can be used to characterize the GR-like region $\mathcal{N}$ of the space-time manifold $\mathcal{M}$.

At this stage, we want to mark the difference of our approach with previous field theoretic directions:\begin{enumerate}
    \item we do not base our model on quantum perturbations. Our cosmological constant does not play the role, as in the traditional and naive understanding, of zeroth term in vacuum energy expansion, and also does not suffer from the orders of magnitude difference with observed values of this approach.
    \item our space-time manifold $\mathcal{M}$ necessarily has a non-trivial boundary $\partial \mathcal{M}$ and is simply connected, in agreement with current observational expectations on topology and in contrast with the non-trivial topology approaches \cite[e.g.][]{Bento:2005un,Bento:2006fm}.
    \item our cosmological constant is the source of topology, for our space-time manifold, as well as a way to ensure \textit{minimum topological requirements} such as \textit{smoothness, simple connected-ness and orientation}, while emerging gravity approaches such as Refs~\citep{Brandenberger:2008nx,LevasseurPerreault:2011mw,Brandenberger:2012um,Afshordi:2014cia} build topology from iterative processes.
\end{enumerate}


\section{Effective topological cosmological constant}\label{sec:EffTCC}

We can calculate the topological numbers for the Pontryagin and Nieh-Yan forms and obtain:
\begin{align}\label{nPnN}
n_P := \mathfrak{Re} \int_{\mathcal{M}} C_P = 0 \;\;\;\;\; ; \;\;\;\;\; n_N :=  \mathfrak{Re} \int_{\mathcal{M}} C_N = 0 \;\;\;\; .  
\end{align}
At the same time, the Euler number:
\begin{align}\label{nEbody}
\mathbb{Z} \;\; \ni \;\; n_E :=  \mathfrak{Re} \int_{\mathcal{M}} C_E = - \frac{16 \Lambda^2}{3 \left( 4 \pi \right)^2} \left\| u^2_{\hat{\varphi}_N} \right\|^2_{L^2} \;\; .
\end{align}
turns out to be finite by the topology of the space-time considered \cite[see][Ch. 11]{nakahara} \cite[and][Secs.~7.22 - 7.26]{nash}. The Euler number $n_E$ can also be written using its representation as an alternate series of the Betti numbers $b_i \left( \mathcal{M}  \right)$ ($i=0,...,4$) \cite[see][Ch. 1]{donaldson}:
\begin{align*}
n_E = \sum_{j=0}^4 \left(-1 \right)^j b_j \left( \mathcal{M} \right) \quad.    
\end{align*}
Poincar\'{e} duality, i.e $b_i = b_{4-i}$ \cite[see][Ch. 1]{donaldson}, can also be used to write:
\begin{align}
\label{nEtop} n_E & = 2 b_0 - 2 b_3 + b_2 = - 2 k^2_E b_3 \;\;.
\end{align}
In Eq.~(\ref{nEtop}), the $i$-th Betti number can be interpreted physically as measuring the number of $i$-th punctures in a manifold. As $\mathcal{M}$ (and consequently $\mathcal{N}$) is assumed to be simply connected, i.e. the manifold is composed by only one connected component, therefore $b_0=1$. Since from Eq.~(\ref{nPnN}) $n_P$ turns out to be null in $\mathcal{M}$, the Hirzebruch signature theorem implies that $b_2 = 2 b$, where $b^+=b^-=b$ and $b^+$, $b^-$ are the dimensions of maximal positive $\mathcal{H}^+$ and negative $\mathcal{H}^-$ subspaces for the form in $H^2 \left( \mathcal{M} ; \mathbb{R} \right) = \mathcal{H}^+ \oplus \mathcal{H}^-$, respectively.

We note that there seems to be no clear evidence of physical objects that can be interpreted as strict $2$-punctures in our Universe since that would appear, for instance, as a naked line of singularities. The cosmic censorship conjecture \cite{Penrose:1969pc} would prescribe it to be zero. However, we do not discard their presence but assume their associated number $b$ to be finite.

Finally, we also follow the cosmic censorship conjecture, stating that in our observable Universe, causality effectively defines horizon 3-hyper-surfaces around space-time singularities in $\mathcal{M}$. Those hyper-surfaces effectively act as $3$-punctures of $\mathcal{M}$, which total number is $b_3$. The clearest example of these singularities are BHs, which observations indicate there should be a large number of. Therefore we assume $b_3$ to be very large compared with $b$. This justifies our factoring in the last line of Eq.~(\ref{nEtop}) with $k_E$ verifying:
\begin{align}\label{kE}
\mathbb{R} \ni k^2_E := 1 - \frac{1 + b}{ b_3}  \lesssim 1 \quad .
\end{align}

The theoretical result thus follows from  Eq.~(\ref{nEbody}) :
\begin{align}\label{eq:Topolambdabody}
\Lambda_{\scriptscriptstyle T}^2 = \frac{ 3 \left( 4 \pi \right)^2 k^2_E b_3 }{8 \left\| u^2_{\hat{\varphi}_N} \right\|^2_{L^2}} \;\;\;\;\;  \Longrightarrow  \;\;\;\;\; \Lambda_{\scriptscriptstyle T} \approx  \frac{ 2 \pi \, k_E \, \mathcal{C} }{ \left\langle \frac{2}{3} \text{Vol} \left( \partial \mathcal{N} \right) \right\rangle^{\frac{1}{2}} } \;\; ,
\end{align}
where we have here used the estimate:
\begin{align}\label{uestimatebody} 
0 \leq \left\| u^2_{\hat{\varphi}_N} \right\|^2_{L^2} \simeq \frac{\text{Vol} \left( \partial \mathcal{N} \right)}{\mathcal{C}^2} < \infty \;\; ,
\end{align}
with $\mathcal{C}^2$ is the Cheeger or isoperimetric constant \cite{Benson,ariascastro} with dimensions of $\left[ \texttt{length} \right]$, and we have defined the average boundary volume per 3-puncture, i.e
\begin{align}
    \frac{\text{Vol} \left( \partial \mathcal{N} \right)}{b_3} := \left\langle \text{Vol} \left( \partial \mathcal{N} \right) \right\rangle \simeq \left\langle \text{Vol} \left( \partial \mathcal{M} \right) \right\rangle ,
\end{align}
where the last line comes from assuming that the content of $\mathcal{N}$ in 3-punctures is typical of that of $\mathcal{M}$.

\section{$\Lambda_{\scriptscriptstyle T}$ Evaluation from black holes}\label{sec:BHcc}

The topological cosmological constant can be evaluated from interpreting expression (\ref{eq:Topolambdabody}). We decompose this evaluation into three main elements that needs separate estimations: the value of the constants $k_E$, $\mathcal{C}$ and of the average spacetime boundary volume $\langle \text{Vol} \left( \partial \mathcal{M} \right) \rangle$.

The ratio of the Euler number to the number of three punctures, $k_E=\sqrt{-\frac{n_E}{2b_3}}$, seen in E.q~(\ref{nEtop}, is considered as the ratio of the topological Euler number to the number of black holes (BHs) contained in the manifold. Indeed as we can carve out the causally disconnected BH event horizons interiors from the rest of spacetime, we consider its boundary to consist in the sum of those hypersurfaces, and thus to correspond to the 3-punctures measured by the third Betti number ($b_3=\mathcal{N}_{BH}$). For any sensible spacetime, $b_3$ should dominate the other terms in Eq.~(\ref{nEtop}), so we can evaluate $k_E\approx 1$.

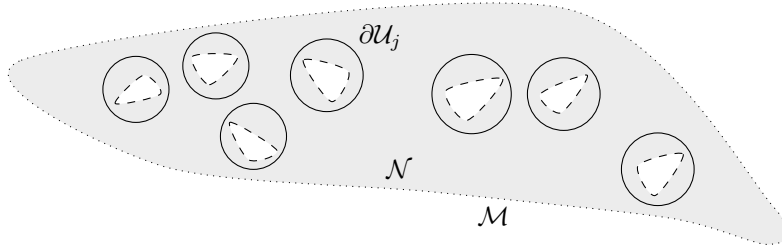
\begin{figure}
\centering
\begin{tikzpicture}[scale=1.25]
\draw [dotted, fill=gray!15] plot [smooth cycle] %
    coordinates {(-2.04, 1.54) (-3.52, 2.66) (2.22, 3.22) %
    (4.48, 1.1)(4.44, 0.7) (3.38, 0.98) (0.84, 1.26)};
\draw [dashed, fill=white!15] plot [smooth cycle] %
    coordinates {(3.04, 1.54) (3.52, 1.66) (3.22, 1.22)};
\draw [dashed, fill=white!15] plot [smooth cycle] %
    coordinates {(2, 2.3) (2.5, 2.5) (2.22, 2.1)};
\draw [dashed, fill=white!15] plot [smooth cycle] %
    coordinates {(-1.7, 2.7) (-1.52, 2.4) (-1.22, 2.7)}; 
\draw [dashed, fill=white!15] plot [smooth cycle] %
    coordinates {(-0.04, 2.54) (-0.52, 2.66) (-0.22, 2.22)};
\draw [dashed, fill=white!15] plot [smooth cycle] %
    coordinates {(1, 2.34) (1.6, 2.46) (1.22, 2.02)};    
\draw [dashed, fill=white!15] plot [smooth cycle] %
    coordinates {(-2.04, 2.3) (-2.52, 2.2) (-2.22, 2.5)}; 
\draw [dashed, fill=white!15] plot [smooth cycle] %
    coordinates {(-1.1, 1.6) (-1.3, 2) (-0.8, 1.7)};
\draw (2.25,2.3) circle (11pt);
\draw (1.27,2.3) circle (12pt);
\draw (-0.27,2.5) circle (11pt);
\draw (-1.45,2.6) circle (10pt);
\draw (-2.3,2.35) circle (10pt);
\draw (-1.05,1.85) circle (10pt);
\draw (3.25,1.5) circle (11pt);
\draw (1.5,1) node {$\mathcal{M}$}; 
\draw (0.5,1.5) node {$\mathcal{N}$}; 
\draw (0.3,2.9) node {$\partial \mathcal{U}_j$}; 
\end{tikzpicture}
\caption{\label{symplified} A simplified two-dimensional version of the space-time manifold $\mathcal{M}$. The manifold $\mathcal{N} \simeq \mathcal{M} \setminus \bigsqcup_j \mathcal{U}_j$ defines the GR regions. The boundary of the manifold is internal and defined by the sum of the boundaries of $\mathcal{U}_j$, i.e $\partial \mathcal{M} \simeq \bigsqcup_j \partial \mathcal{U}_j$.}
\end{figure}

The latter procedure effectively defines a non-trivial manifold boundary $\partial \mathcal{M}$ even if the volume of the manifold $\text{Vol} \left( \mathcal{M} \right)$ is taken to be infinite. We can exemplify this picture by thinking of a simplified two-dimensional version shown in  Figure. \ref{symplified}.

The isoperimetric or Cheeger constant, $\mathcal{C}$, gives the ratio of the spacetime $4$-volume to its boundary hypersurface $3$-volume. Although it is generally unknown, it was calculated in some specific cases \cite{ASENS, Hoffman}, and is thus considered to be of order $\sim 10^1$.

Once we have identified the boundary of the spacetime to the horizons of its BHs, the evaluation of the topological cosmological constant relies on the estimation of the average spacetime BH horizon volume $\langle \text{Vol} \left( \partial \mathcal{M} \right) \rangle$. 
This requires to make some assumptions.

\subsection{Average BH boundary volume estimation}

To compute the volume of the average BH in our Universe, the following assumptions are made: 
\begin{enumerate*}[label=(\alph*)]
    \item as the Universe's total boundary is taken to be the sum of all BHs horizons, that hypersurface is assumed to be given by BHs equivalent Schwarzschild horizons. This neglects Kerr Horizons deformations and the different horizons shapes taken at the moment of BH mergers, assuming each BH can be approximated by an isolated Schwarzschild horizon.
    \item the Universe's BH distribution is assumed to be represented by our past lightcone BHs observations and present knowledge of that distribution is sufficient for the calculation, giving the boundary volume as its first moment. 
    \item the distribution's average volume per BH in the Universe is well approximated by the volume of a BH starting with the average BH mass.
    \item the volume of a BH of given initial mass is approximated to proceed from an almost instant creation with a mass picked in the observed lightcone distribution followed by a long Hawking radiation phase evaporation.
\end{enumerate*}

The computation of our Universe's average BH volume requires to evaluate, from observed BH distribution, the average BH mass of the Universe, to get the BH horizon volume of a fixed given mass, so as to put them together in an evaluation of the Universe's boundary volume
\subsubsection{Average BH mass evaluation}
We chose to evaluate the BH mass distribution \Jav{estimated from }
the observations and computations of Refs~\cite{Kovetz:2016kpi,Mutlu-Pakdil2016,Garcia-Bellido:2017fdg,Christian:2018mjv}. 

\begin{table}\Jav{
\begin{tabular}{c|c|c}
\hline 
 & average mass/sdt dev.  & references \tabularnewline
& (in solar mass units $M_{\odot}$) & \tabularnewline
\hline
Stellar BH & $M_{S}=10^{1.02_{-0.41}^{+0.21}}$ & \cite{Christian:2018mjv}\cite[Figs. 2, 6]{Kovetz:2016kpi} \tabularnewline
\hline
Primordial BH & $M_{P}=10^{1.96_{-0.21}^{+0.14}}$ & \cite[using Fig 7]{Garcia-Bellido:2017fdg} \tabularnewline
\hline
Intermediate Mass BH & $M_{I}=10^{4.19_{-0.21}^{+0.14}}$ & \cite[using Fig 7]{Garcia-Bellido:2017fdg} \tabularnewline
\hline
Super Massive BH & $M_{SMBH}=10^{11.69_{-0.39}^{+0.20}}$ & \cite[Fig. 10]{Mutlu-Pakdil2016}\tabularnewline
\hline 
\end{tabular}
\caption{\label{tab:BHpeaks}Estimation of the first moments for each range BH mass distributions.}}
\end{table}

From them, we have extracted averages and standard deviations from the expected peaks in the BH distributions around stellar mass BHs, Primordial BHs (PBHs), Intermediate mass BHs (IMBHs) and Super Massive BHs (SMBHs)\Jav{ (see table \ref{tab:BHpeaks} for  intermediate evaluations). Given that the data is of logarithmic nature, we estimated 
the average mass using}
a geometrical weighted average\Jav{\footnote{\Jav{Estimating the weights from Ref.~\cite[using Fig 7]{Garcia-Bellido:2017fdg}, we used $\langle M \rangle=\left(M_S^{1+2k} M_P^{1+5k} M_I M_{SM}^{1+k}\right)^\frac{1}{4+8k}$, with $k$ given by the estimation of difference between} \Jav{the $I$ and $SM$ peaks and calculated within the range $\left[k_1=\frac{3}{10}..k_2=\frac{1}{3}\right]$, to yield $\langle M \rangle= \left(M_S^{1.63} M_P^{2.58} M_I M_{SM}^{1.32}\right)^\frac{1}{6.53}$. This final result is somewhat unaffected by the choosing of the different $k$s since their effects are minimized by the geometric mean.}}}.
\Jav{We emphasize that this is an estimation based on available knowledge and does not claim to reflect a rigorous measurement.} \Mov{Incidentally, the use of the geometric mean also minimizes the effects of the uncertainties in the data regarding black hole mass distribution ranges which still remain somewhat unexplored or lacking observational data. All of these, combined with a conservative treatment of errors allows us to obtain:}
\begin{align}
    \left\langle M_{BH}\right\rangle \sim & 10^{4.04^{+0.49}_{-0.61}}M_\odot \;\;, \label{eq:AvgBHmassU}
\end{align}
which is slightly below our evaluation of the IMBH average and above the dominating PBH peak average estimate. \Mov{This value is expected to be improved by new observational data in the future.}

\subsubsection{BH volume of a given mass}

For a given BH mass $M$, the BH formation phase is neglected since its dynamical time is expected to be considerably much less than the Hawking radiation evaporation time. The horizon volume is therefore estimated  considering the BH appears at creation with initial mass $M$ and evaporates through Hawking radiation, until the complete BH evaporation, evaluated considering the mass-energy loss  \cite[e.g.][]{Wald:1984rg,Carroll:2004st}. As previously mentioned, we consider that mass to be ascribed to a simple isolated Schwarzschild BH. The calculation of such volume yields
\begin{align}
     \text{Vol}_{BH} \left( M \right)=&1.96\times 10^{87}\left(\frac{M}{M_\odot}\right)^5 m^3 \;\;.\label{eq:BHBounVol}
\end{align}
\subsubsection{Universe's total boundary volume}
Following our assumptions above, the resulting boundary volume of the Universe can be evaluated by the product of the total number of BHs with the average BH volume, given by introducing the average BH mass estimate (\ref{eq:AvgBHmassU}) into the BH boundary volume evaluation (\ref{eq:BHBounVol}), so we obtain 
\begin{align}
    \text{Vol} \left( \partial \mathcal{M} \right) \sim &  \mathcal{N}_{BH}10^{107.5^{+2.5}_{-3.1}} m^3 \;\;.\label{eq:BounVol}
\end{align}
\subsection{Evaluating the topological cosmological constant}
Now that we have the estimate of the BH boundary volume (\ref{eq:BounVol}), we can finally input it into the model result  (\ref{eq:Topolambdabody}) to get $\Lambda_{\scriptscriptstyle T}$ as a function of the Cheeger and $k_E$ constants 
\begin{align}
\Lambda_{\scriptscriptstyle T} \approx 10^{-52.9_{-1.3}^{+1.5}} \, k_E \mathcal{C} \approx 10^{-52.9_{-1.3}^{+1.5}} \mathcal{C} \;\; ,\label{eq:TopoLambdaWithC}
\end{align}
since we previously argued that $k_E\approx1$, only the Cheeger constant remains.

\subsubsection{Can $\Lambda_{\scriptscriptstyle T}$ be the observed cosmological constant?}

Converting the latest Planck observations \cite{Planck2018} in the appropriate units, we compute $\Lambda_{\scriptscriptstyle O}=10^{-51.08\pm0.01}m^{-2}$. The isoperimetric constant has been evaluated for a 4-manifold with null sectional curvature, Ref.~\cite{ASENS, Hoffman} and the value  $\mathcal{C} = 11.8$ was obtained. Approximating the Universe's value with it,
\begin{align}
 \Lambda_{\scriptscriptstyle T} \approx 10^{-51.8^{+1.5}_{-1.3}} m^{-2} \;\;,   
\end{align}
and thus our $\Lambda_{\scriptscriptstyle T}$ estimate is compatible with the observed $\Lambda_{\scriptscriptstyle O}$. Given that $\mathcal{C} \sim O(10)$, we argue that the topology of the Universe is a serious candidate for the cosmological constant origin, and this giving naturally its low value and avoiding the cosmological constant fine tuning problem.
\subsubsection{The Universe isoperimetric constant can be measured}
Although the previous evaluation gives the correct observed cosmological constant, the actual value of $\mathcal{C}$ for our Universe remains undetermined. If we assume $\Lambda_{\scriptscriptstyle T}=\Lambda_{\scriptscriptstyle O}$, Eq.~(\ref{eq:TopoLambdaWithC}) can be used to measure, through the average BH mass and volume estimates of this work and the current cosmological constant observations \cite{Planck2018}, the value of the Universe's isoperimetric constant which contributes to the determination of the topology of the Universe. We obtain
\begin{align}
    \mathcal{C}=10^{1.82^{+1.31}_{-1.51}},
\end{align}
which gives a reasonable value compared to expectations.

\section{Conclusions}\label{sec:Conclusion}

Introducing the topological invariants \cite{donaldson,Sengupta,Kaul,Baekler,Dyer,Lorca} in the gravitation theory as Lagrange multipliers induces an effective cosmological constant
. The extra degrees of freedom and restrictions of the topological invariants in an Einstein-Cartan gravity \cite{Dona,Giulini,nakahara,hatcher} are handled by a reasonable ansatz. They allow to obtain a GR-like behaviour dynamically, driven by the invariants coupling zero forms that can be considered as effective dark energy fields. Indeed, we argue that their dynamics constrain the value of the expansion acceleration to agree with the BH boundary of the Universe. The induced effective topological cosmological constant in the GR-like theory is produced by the Euler invariant. It depends on the Betti numbers of the spacetime \cite{milnor,donaldson}, the isoperimetric constant \cite{Benson} and the volume of the manifold boundary. The interpretations of $\Lambda_{\scriptscriptstyle T}$ in terms of number and surface of BHs in the Universe allows to evaluate it from current estimates of the BH distribution of the Universe \cite{Kovetz:2016kpi,Mutlu-Pakdil2016,Garcia-Bellido:2017fdg,Christian:2018mjv}. We found, with some reasonable assumptions on the isoperimetric constant, that it is compatible with current cosmological constant observations \citep{Hinshaw:2012aka,Bennett:2012zja,Aghanim:2012vda,Ade:2013sjv,Planck2018}, escaping the cosmological constant fine tuning problem \citep{Weinberg:1988cp,Weinberg:2008zzc}. Our BH volume evaluation being based on some BH distribution estimations that rely on gravitational waves and BH population knowledge, future improvements in those domains, both experimental and theoretical, are expected to allow narrowing on the topological cosmological constant estimation, increasing the testability of the approach compared with the Universe's acceleration or geometrical observations of $\Lambda_{\scriptscriptstyle O}$. 
We have here ignored the behaviour of our results in a quantum setting, and in particular how  the cosmological constant (\ref{eq:TopoLambdaWithC}) remains robustly unaffected by  quantum fluctuations. However these constitute interesting questions and a thorough study will be tackled in future works.
We conjecture that the remaining unknown isoperimetric constant could be independently obtained from the development of an emerging geometry approach \citep{Brandenberger:2008nx,LevasseurPerreault:2011mw,Brandenberger:2012um,Afshordi:2014cia} to the dynamic theory of the manifold topology, with the potential to perhaps clarify the cosmological constant coincidence problem \citep{Amendola:1999er,TocchiniValentini:2001ty,Zimdahl:2001ar,Zimdahl:2002zb} 
from topological considerations as well as current estimates of the BH distribution of the Universe \cite{Kovetz:2016kpi,Mutlu-Pakdil2016,Garcia-Bellido:2017fdg,Christian:2018mjv}. We found, with some reasonable assumptions on the isoperimetric constant, that it is compatible with current cosmological constant observations \citep{Hinshaw:2012aka,Bennett:2012zja,Aghanim:2012vda,Ade:2013sjv,Planck2018}, escaping the cosmological constant fine tuning problem \citep{Weinberg:1988cp,Weinberg:2008zzc}. Our BH volume evaluation being based on some BH distribution estimations that rely on gravitational waves and BH population knowledge, future improvements in those domains, both experimental and theoretical, are expected to allow narrowing on the topological cosmological constant estimation, increasing the testability of the approach compared with the Universe's acceleration or geometrical observations of $\Lambda_{\scriptscriptstyle O}$. We conjecture that the remaining unknown isoperimetric constant could be independently obtained from the development of an emerging geometry approach \citep{Brandenberger:2008nx,LevasseurPerreault:2011mw,Brandenberger:2012um,Afshordi:2014cia} to the dynamic theory of the manifold topology, with the potential to perhaps clarify the cosmological constant coincidence problem \citep{Amendola:1999er,TocchiniValentini:2001ty,Zimdahl:2001ar,Zimdahl:2002zb} as well from topological considerations.


\section*{Acknowledgements}

The authors wish to thank M.Fontanini and E. Huguet for very useful discussions, as well as O. Bertolami for interesting perspectives.
The work of M.Le~D. has been supported by Lanzhou University starting fund, the Fundamental Research Funds for the
Central Universities (Grant No.lzujbky-2019-25) and PNPD/CAPES20132029. M.Le~D. also wishes to acknowledge IFT/UNESP for hosting the beginning of this project.


\bibliographystyle{elsarticle-num}

\bibliography{sample}






\end{document}